\documentclass[onecolumn,11pt]{article}
\usepackage[top=1in, bottom=1in, left=1.25in, right=1.25in]{geometry}

\usepackage{amssymb}
\usepackage{amssymb}
\usepackage{amsmath}
\usepackage{amssymb}
\usepackage[amsmath,thmmarks]{ntheorem}
\usepackage{indentfirst}

\usepackage[dvips]{graphicx}

\title{Zero-point attracting projection algorithm for sequential compressive sensing}

\author{Yang You, Jian Jin, Wei~Duan, Ningning~Liu, Yuantao~Gu\thanks{This work is partially supported by National Natural Science Foundation of China (NSFC 60872087 and NSFC U0835003). The corresponding author of this paper is Yuantao Gu (email: gyt@tsinghua.edu.cn)}, and Jian Yang}

\date{received July 27, 2011; accepted Jan. 04, 2012.\\\vspace{1em}
This article appears in \textsl{IEICE Electronics Express}, 9(4):314-319, 2012.}

\begin{document}

\maketitle

\begin{abstract}
Sequential Compressive Sensing, which may be widely used in sensing
devices, is a popular topic of recent research. This paper proposes
an online recovery algorithm for sparse approximation of sequential
compressive sensing. Several techniques including warm start, fast
iteration, and variable step size are adopted in the proposed
algorithm to improve its online performance. Finally, numerical
simulations demonstrate its better performance than the relative
art.

\end{abstract}

\section{Introduction}

Compressive sensing (CS) \cite{CSbase,CS1} is a recently proposed
concept that enables sampling below Nyquist rate, without (or with
little) sacrificing reconstruction quality. Based on exploiting the
signal sparsity in typical domains, CS methods can be used in the
sensing devices, such as MR imaging \cite{MRI} and AD conversion
\cite{ADC}, where the devices have a high cost of acquiring each
additional sample or a high requirement on time. Therefore, as the
sparsity level is often not known a priori, it can be very
challenging to use CS in practical sensing hardware.

Sequential compressive sensing \cite{SCS} can effectively deal with
the above problems. Sequential CS considers a scenario where the
observations can be obtained in sequence, and computations with
observations are performed to decide whether these samples are
enough. Consequently, it is allowed to recover the signal either
exactly or to a given tolerance from the smallest possible number of
observations. There have been several recovery algorithms for
sequential CS. Asif \cite{Asif} solved the problem by homotopy
method. Garrigues \cite{lasso} discussed the Lasso problem with
sequential observations.

This work extends a recent proposed zero-point attracting projection
(ZAP) algorithm \cite{ZAP} to the scenario of sequential CS. ZAP employs an approximate $l_0$ norm as the sparsity constraint and updates in the
solution space. Comparing with the existing algorithms, it needs
fewer measurements and lower computation complexity. Therefore the
new algorithm can provide a much more appropriate solution for
practical sensing devices, which is validated by numerical
simulations.

\section{Background}

\subsection{Compressive sensing}

Suppose $\bf x$ is an unknown sparse signal, which is $N$-length but
has only $K$ nonzero entries, where $K\ll N$. In their ice-breaking
contributions, Candes et al suggested to measure $\bf x$ with
under-determined observations, i.e. $\bf y=Ax$, where $\bf A$
consists of random entries and has much fewer rows than columns.
They also proved that $l_0$ norm or $l_1$ norm constraint
optimization can successfully recover the unknown signal with
overwhelming probability,
\begin{equation}\label{problem}
    \hat{\bf x} = \arg\min_{\bf x}\|{\bf x}\|_{0\,{\rm or}\,1}, \,{\rm subject\ to}\ {\bf y}={\bf A}{\bf x}.
\end{equation}
There are many methods proposed to solve (\ref{problem}), of which
concerned in this work is ZAP.

\subsection{Zero-point attracting projection}

ZAP iteratively searches the sparsest result in solution space. The
recursion starts from the least square optimal solution, ${\bf x}(0)
= {\bf A}^{\dagger}{\bf y}$, where ${\bf A}^{\dagger}={\bf A}^{\rm
T}({\bf A}{\bf A}^{\rm T})^{-1}$ denotes the Pseudo-inverse matrix
of ${\bf A}$. In the $n$th iteration, the solution is first updated
along the negative gradient direction of a sparse penalty,
\begin{equation}\label{ZA}
    \tilde{\bf x}(n+1)={\bf x}(n)-\kappa\cdot\nabla g({\bf x}(n)),
\end{equation}
where $g(\cdot)$ denotes a sparse constraint function and $\kappa$
denotes the step size. In the reference, an approximate $l_0$ norm
is employed and the corresponding $i$th entry of $\nabla{ g}(\cdot)$
is
\begin{equation}\label{l0Penalty}
    \{\nabla g\}_i({\bf x})=\left\{
    \begin{array}{cl} \alpha \cdot{\rm sgn}(x_i)-{\alpha^2}x_i & |x_i|\leq\frac{\textstyle1}{\textstyle\alpha}; \\
    0 & {\rm elsewhere}, \end{array}\right.
\end{equation}
where $\alpha$ is a controlling parameter and it is readily
recognized that the penalty tends to $l_0$ norm as $\alpha$
approaches to infinity. Then $\tilde{\bf x}(n+1)$ is projected back
to the solution space to satisfy the observation constraint,
\begin{equation}\label{Projection}
    {\bf x}(n+1)={\bf P}\tilde{\bf x}(n+1)+{\bf Q},
\end{equation}
where $\bf P=I-A^{\dagger}A$ is defined as projection matrix and
$\bf Q=A^{\dagger}y$. Equation (\ref{ZA}) appears that an attractor
locates at the zero-point is pulling the iterative solution to be
sparser, as explains the first part of the algorithm's name. The
last part comes from (\ref{Projection}), which means that
$\tilde{\bf x}(n+1)$ is projected back to the solution space.


\subsection{Sequential compressive sensing}

Imagining a scenario that the samples are measured in realtime. At
time $m$, an $m$-length measurement vector ${\bf y}_m={\bf A}_m{\bf
x}$ is collected and utilized to solve the sparsest solution by
(\ref{problem}). If the available measurements are not enough to
recover the original sparse signal, a new sample $y_{m+1}={\bf
a}_{m+1}^{\rm T}{\bf x}$ is generated at time $m+1$, where ${\bf
a}_{m+1}$ denotes the sampling weight vector. Thus the problem
becomes solving ${\bf y}_{m+1}={\bf A}_{m+1}{\bf x}$, where
$$
    {\bf y}_{m+1} = \left[{\bf y}_m \atop y_{m+1}\right],\qquad {\bf A}_{m+1}=\left[{\bf A}_m \atop {\bf a}_{m+1}^{\rm T}\right].
$$
Obviously, it is a waste of resources if the recovery algorithm is
re-initialized without the utilization of earlier estimate, i.e. the
available result at time $m$. Consequently, the basic aim of
sequential compressive signal reconstruction is to find an effective
method of refining $\hat{\bf x}_{m+1}$ based on the information of
$\hat{\bf x}_{m}$.

\section{Online ZAP for sequential compressive sensing}

For conciseness, the detailed iteration procedure of online ZAP is
provided in Tab.\ref{Tab1}. It can be seen that online ZAP has two
recursions. The inner iteration is to update the solution by ZAP
with the given measurements. The outer iteration is for sequential
input. In order to improve the performance, several techniques are
used in online ZAP and they are discussed in the following
subsections.

\begin{table}[ht]
\begin{center}
\caption{The Procedure of Online ZAP}\label{Tab1}
\begin{small}
\begin{tabular}{l}
\hline
{\bf Input:} $\alpha,\kappa_0,\eta_1,\eta_2,T,Q;$\\
{\bf Initialize online ZAP:} ${\bf x}_0(0)=0, m=0, n_{0} = 0$.\\
{\bf Repeat:} (for time instant $m$); \\
~~~~for the new ${\bf a}_{m+1}$ and $y_{m+1}$,
calculate ${\bf\Gamma}_{m+1}$ by (\ref{updategamma}) \\ ~~~~~~~~and then produce ${\bf P}_{m+1}$ and ${\bf Q}_{m+1}$;\\
~~~~decrease step size by $\kappa_{m+1}=\eta_1\kappa_m$;\\
~~~~$m=m+1$;\\
~~~~{\bf Initialize ZAP:} $n=0, {\bf x}_{m}(0) = {\bf x}_{m-1}(n_{m-1}), \kappa=\kappa_m$;\\
~~~~{\bf Repeat:} (for the $n$th iteration of ZAP);\\
~~~~~~~~Update $\tilde{\bf x}_m(n+1)$ with the zero attraction by (\ref{ZA}) and (\ref{l0Penalty});\\
~~~~~~~~Project ${\bf x}_{m}(n+1)$ back to the solution space by (\ref{Projection});\\
~~~~~~~~if $\|{\bf x}_{m}(n+1)\|_1 > \|{\bf x}_{m}(n)\|_1$ and $\kappa>\kappa_m/Q$, \\
~~~~~~~~~~~~decrease the step size by $\kappa
=\eta_2\kappa$;\\
~~~~~~~~$n=n+1$;\\
~~~~{\bf Until:} $n>T$ or $\kappa<\kappa_m/Q$. \\
~~~~$n_m = n-1$;\\
{\bf Until:} online ZAP stop criterion is satisfied.\\
 \hline
\end{tabular}
\end{small}
\end{center}
\end{table}

\subsection{Warm Start}

ZAP works in an iterative way to produce a sparse solution via
recursion in the solution space. In the online scenario, the
previous estimate can be used to initialize the incoming iteration,
i.e. ${\bf x}_{m+1}(0) = {\bf x}_m(n_m)$, where $n_m$ denotes the
maximum iteration number at time $m$.

\subsection{Fast Iteration}

The Pseudo-inverse matrix ${\bf A}^{\dagger}$ plays an important
role in the recursion of (\ref{Projection}). Considering the high
computational cost of matrix inverse operation, ${\bf A}^{\dagger}$
is generally prepared before iterations. However, in the online
scenario, ${\bf A}_m$ becomes time-dependent and ${\bf
A}_m^{\dagger} = {\bf A}_m^{\rm T}\left({\bf A}_m{\bf A}_m^{\rm
T}\right)^{-1}$ need to be recalculated in each time instant. In
order to reduce the complexity, the Pseudo-inverse matrix is updated
iteratively.

Define ${\bf\Gamma}_m = \left({\bf A}_m{\bf A}_m^{\rm
T}\right)^{-1}$, which is already available after time $m$.
Consequently, as the new sample is arriving, using basic algebra one
has the recursion
\begin{align}
    {\bf\Gamma}_{m+1} &=  \left[\left[{\bf A}_m \atop {\bf a}_{m+1}^{\rm T}\right]\left[{\bf A}_m^{\rm T} \; {\bf a}_{m+1}\right]\right]^{-1}
    = \left[\begin{matrix}{\bf\Gamma}_{m}^{-1} & {\boldsymbol\alpha}_m \\ {\boldsymbol\alpha}_m^{\rm T} & \beta_m\end{matrix}\right]^{-1}\nonumber\\
    &= \left[\begin{matrix}{\bf\Gamma}_m+\theta_m{\bf\Gamma}_m
   {\boldsymbol\alpha}_m{\boldsymbol\alpha}_m^{\rm T}{\bf\Gamma}_m & -\theta_m{\bf\Gamma}_m{\boldsymbol\alpha}_m\\
   -\theta_m{\boldsymbol\alpha}_m^{\rm T}{\bf\Gamma}_m & \theta_m\end{matrix}\right],\label{updategamma}
\end{align}
where
$$
    {\boldsymbol\alpha}_m={\bf A}_m {\bf a}_{m+1},\qquad
    \beta_m ={\bf a}_{m+1}^{\rm T}{\bf a}_{m+1},\qquad
    \theta_m=\frac{1}{\beta_m-{\boldsymbol\alpha}_m^{\rm T}
    {\bf\Gamma}_m{\boldsymbol\alpha}_m}.
$$


%

\subsection{Variable step size}


As the step size in gradient descent iterations, the parameter
$\kappa$ controls a tradeoff between the speed of convergence and
the accuracy of the solution. In order to improve the performances
of the proposed algorithm, the idea of variable step size is taken
into consideration. The control scheme is rather direct: $\kappa$ is
initialized to be a large value after new sample arrived, and
reduced by a factor as long as the iteration is convergent. The
reduction is repeated several times until $\kappa$ is sufficiently
small. Since the algorithm has two recursions, we employ $\eta_1$
and $\eta_2$ to denote the decreasing speed of outer and inner
iteration, respectively. In addition, $\kappa$ is no longer decreased when the step size is rather small.

\subsection{Stop Rules}

There are two kinds of recursions requiring stop rules in the online
ZAP algorithm. Firstly, after the $m$th sample arrived, ${\bf
x}_m(n)$ iterates with $n$ to produce the best estimate based on the
$m$ measurements. The inner iteration should stop after the
algorithm reaches steady state, which means the sparsity penalty
starts increasing. Consequently, the inner recursion stops (a) when
the number of reductions of $\kappa$ reaches one-$Q$th of its
initial value or (b) when the number of iterations reaches the bound
$T$.

Secondly, as soon as the sparse signal is successfully
reconstructed, the following samples are no longer necessary and the
sensing procedure stops. Therefore, the outer recursion stops when
the estimate error is below a particular value $\varepsilon$.

\begin{figure}
\centering
\includegraphics[width=4in]{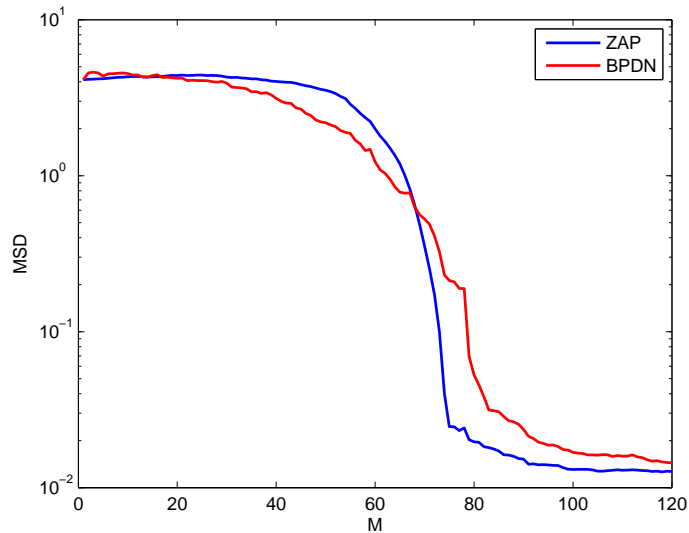}
\caption{The reconstruction MSD versus the number of measurements $M$.} \label{varm}
\end{figure}

\section{Experiment and Discussion}
Computer simulation is presented in this section to verify the
performance of the proposed algorithm compared with typical
sequential CS reconstruction algorithm for solving BPDN problem
\cite{Asif}, whose MATLAB code can be downloaded from the website
\cite{website}. In the following experiment, the entries of each row
of $\bf A$ are independently generated from normal distribution. The
locations of $K$ nonzero coefficients of sparse signal $\bf x$ are
randomly chosen with uniform distribution $[1,N]$. The corresponding
nonzero coefficients are Gaussian with mean zero and unit variance.
The system parameters are $N=256$ and $K=20$. The number of
measurements $M$ increases form $1$ to $120$. The parameters for
BPDN are set as the recommended values by the author. The parameters
for online ZAP are $\alpha=1$, $\kappa_0=0.02$, $\eta_1=0.99$,
$\eta_2=0.8$, $T=50$, $Q=2000$. The simulation is repeated ten
times, then Mean Square Derivation (MSD) between the original signal
and reconstruction signal as well as the average running time
calculated.

\begin{figure}
\centering
\includegraphics[width=4in]{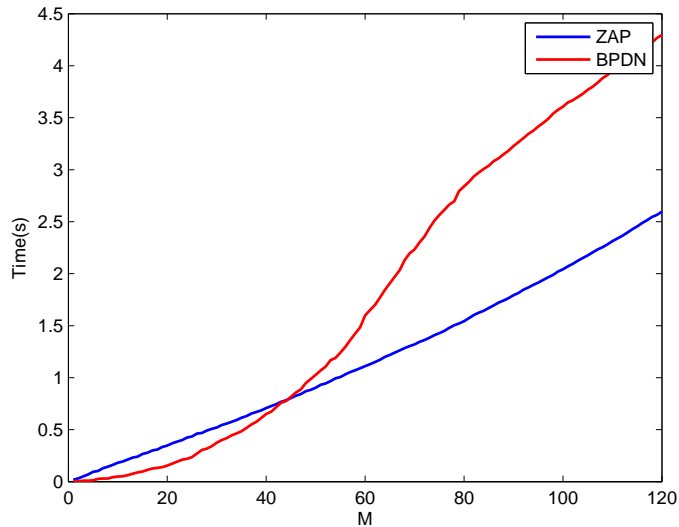}
\caption{The CPU running time versus the number of measurements $M$.} \label{time}
\end{figure}

Figure \ref{varm} shows MSD curve according to $M$. As can be seen,
the performance of ZAP is better than that of BPDN. When the sparse
signal is recovered successfully, the number of measurements BPDN
needs is larger than $80$, while the number ZAP algorithm needs is
less than 80. Figure \ref{time} demonstrates the CPU running time as
$M$ increases. Again, ZAP has the better performance. The CPU time
of BPDN is twice than that of ZAP for successful recovery (according
to Fig.\ref{varm}, here $M$ is chosen as $80$ for comparison).

\section{Conclusion}
We have introduced in this paper a new online signal reconstruction
algorithm for sequential compressive sensing. The proposed algorithm
extends ZAP to sequential scenario. And in order to improve the
performance, some methods, including the warm start and variable
step size, are adopted. The final experiment indicates that the
proposed algorithm needs less measurements and less CPU time than
the reference algorithm.

\end{document}